\documentclass[
amsmath,amssymb,
twocolumn
 ]{revtex4}
\usepackage{graphicx}% Include figure files
\usepackage{dcolumn}% Align table columns on decimal point
\usepackage{bm}% bold math
\usepackage{hyperref}% add hypertext capabilities
\usepackage{bbold}
\usepackage{color}
\renewcommand{\d}[2]{\ensuremath{\frac{\text{d} #1}{\text{d} #2}}}

\newcommand{\pd}[2]{\ensuremath{\frac{\partial #1}{\partial #2}}}

\newcommand{\ket}[1]{\ensuremath{\left| #1 \right>}}

\renewcommand{\exp}[1]{\ensuremath{ \; \text{exp} \left( #1 \right) } }

\begin{document}

\preprint{APS/123-QED}

\title{Quantum dot at a Luttinger liquid edge}% Force line breaks with \\

\author{Colin Rylands} 
\email{rylands@physics.rutgers.edu}
\author{Natan Andrei}
\email{natan@physics.rutgers.edu}
\affiliation{Department of Physics, Rutgers University
Piscataway, New Jersey 08854.
}

\date{\today}% It is always \today, today,
              
 \begin{abstract}
We study a system consisting of a Luttinger liquid coupled to a quantum dot on the boundary. The Luttinger liquid is expressed in terms of fermions interacting via density-density coupling  and the dot is modeled as an interacting resonant level on to which the bulk fermions can tunnel. We solve the Hamiltonian exactly and construct all eigenstates.  We study both the zero and finite temperature properties  of the system, in particular we compute the exact dot occupation as a function of the dot energy in all parameter regimes. The system is seen to flow from weak to strong coupling for all values of the  bulk interaction,  with the flow characterized by a non-perturbative Kondo scale. We identify the critical exponents at the  weak and strong coupling regimes. 
  \end{abstract}

\maketitle
\section{Introduction}

The low energy physics of many one dimensional systems are successfully captured by a Luttinger liquid  description\cite{haldane}. Among these are quantum wires with screened Coulomb interaction and spin chains. The description entails expanding about the two Fermi points so that the system consists of left or right moving particles and  low energy interaction processes. The presence of interaction causes the Fermi surface to be destroyed so that the excitations are collective bosonic density perturbations. 

 The effects of the electrons being dissolved  are  most dramatic when the system is coupled to an impurity and in particular  to a quantum dot  \cite{KF}. Quantum dots are artificial atoms created by confining a two dimensional electron gas to small enough size  that its energy levels become discrete. The potential on the dot and leads can be tuned so that only one of these levels is available to be occupied. 
Therefore when coupled to a Luttinger liquid there exists a competition between the tunneling which is mediated by electrons and the large number of bosons excited as an electron is added to the bulk. 

Luttinger liquids as well as Luttinger impurity systems are readily created in experiments as  carbon nano tubes \cite{LLCNT} \cite{JCNT}, fractional quantum Hall edges \cite{LLRMP}, \cite{FQHLL}, \cite{FQHLL2}, screw dislocations in $^4$He \cite{Boninsegni} and also $^4$He confined to nanopores \cite{DelMae}\cite{Duc}. It can further be realized in a quasi one dimensional setting where the Luttinger liquids describe resistive higher dimensional leads \cite{Uni}\cite{Meb}\cite{qpt}. Most exciting perhaps is the ability to create these systems using cold atom gases \cite{LLcold}\cite{ColdIRL}\cite{Jiang} where the ability to tune interaction strength, hopping parameters combined with precision measurements allows one to explore regions inaccessible to other methods \cite{CA}. 
\begin{figure}
\centering
\includegraphics[trim = 0mm 5mm 0mm 3mm, clip, width=0.4\textwidth]{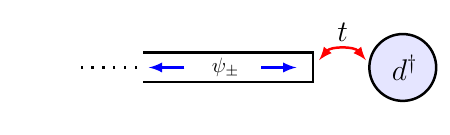}
\caption{Our system consists of a semi infinite Luttinger liquid coupled to a quantum dot modeled as a resonant level. The Luttinger liquid consists of left and right moving interacting fermions which can tunnel to and from the level and experience a Coulomb force from an occupied dot.}
\end{figure}

In this paper we study a spinless Luttinger liquid coupled to a quantum dot at the boundary as depicted in FIG. 1.  The model can describe a quantum dot placed at the end of a spin-polarized nano wire  or  placed in the middle of a fractional quantum Hall edge \cite{LLRMP}. We  solve the model exactly using Bethe Ansatz and identify the ground state and excitations of the model for all parameter regimes. The ground state occupation of the dot is then calculated as a function of the dot energy. From this we determine the renormalization group picture of the system flowing from weak to strong coupling and identify the fixed points as well as the leading relevant and irrelevant operators around them. The finite temperature properties are then studied and the free energy calculated with the same behavior 
found  of a weakly coupled localized dot at high temperature and a  strongly coupled delocalized dot  at low temperatures with the crossover characterized by a dynamically generated energy scale.

\section{The Hamiltonian}
The Hamiltonian of the (spinless) Luttinger liquid  in its fermionic form is given by,
\begin{eqnarray}\nonumber
   H_{\text{LL}}&=&-i  \int_{-L/2}^{0}  dx (\psi^\dagger_+ \partial_x 
\psi_+   -\psi^\dagger_- \partial_x \psi_-)
   \\\label{H}&&+ 4g \int_{-L/2}^{0}   dx  \,\psi_+^\dag(x)\psi^\dag_-(x)\psi_-(x)\psi_+(x)
\end{eqnarray}
where $\psi^\dag_{\pm}$ are right and left moving fermions restricted to the space $x\in[-L/2,0]$ which interact with a point like interaction of strength $g$ \cite{TG}. There are two conserved charges present in $ H_{LL}$ namely the number of left and right movers, $\hat{N}_\pm= \int_{-L/2}^{0}  \psi^{\dagger}_{\pm}(x) \psi_{\pm}(x) $, which are combined to a single conserved charge by the boundary condition $\psi_-(0)=-\psi_+(0)$ mixing the two chirality particles.  

By itself this Hamiltonian is easy to diagonalize and is most conveniently done   by unfolding the system to the full line  using $\psi_-(x)=-\psi(-x)$ and $\psi_+(x)=\psi(x)$ for $x\leq 0$. The result of this is
\begin{eqnarray}\nonumber   H'_{\text{LL}}&=&-i  \int_{-L/2}^{L/2}  dx \psi^\dagger \partial_x 
\psi \\
&&\int_{-L/2}^{L/2} 4g\psi^\dag(x)\psi^\dag(-x)\psi(-x)\psi(x)
\end{eqnarray}
so that only right movers are present but the interaction is now non local  and the system extends from $-L/2$ to $L/2$.  In the unfolded model the interaction occurs between two particles only when $x_1+x_2=0$  and we can expand the wavefunction in plane waves and in regions where $x_1+x_2>0$ or $<0$. The two particle eigenstate with energy $E=k_1+k_2$ is consequently given by  \cite{Thirring}, 
\begin{eqnarray}\nonumber
\ket{k_1,k_2}=\int \!\!\mathrm{d} \vec{x}\left[\theta(x_1+x_2)+e^{i\phi}\theta(-x_1-x_2)\right]\\
\times\prod_{j=1}^2e^{ik_jx_j}\psi^\dag(x_j)\ket{0}
\end{eqnarray}
where $\phi=-2\arctan{(g)}$ is the two particle phase shift and $\theta(x)$ are Heaviside functions. The generalisation to $N$ particles is straightforward and upon doing so the spectrum can be determined by imposing periodic boundary conditions $\psi(L/2)=\psi(-L/2)$ in the unfolded language which corresponds to an open boundary condition, $\psi_+(-L/2)=-\psi_-(-L/2)$ in the folded system. This constrains the single particle energies $k_j,\,j=1,\dots N$ according to 
\begin{eqnarray}\label{LLbae}
e^{-ik_jL}=e^{i(N-1)\phi}
\end{eqnarray} 
with the total energy being the sum of these $E=\sum_j^Nk_j$. Thus the energy levels of a Luttinger liquid in a box are shifted by a constant compared to those of a free model.

The system can likewise be described using bosonization so that the Hamiltonian takes the form \cite{TG},
\begin{eqnarray}\nonumber
H_{LL}= \frac{1}{4 \pi} \int( K (\nabla \varphi)^2+   \frac{1}{K}\Pi^2) \,dx
 \end{eqnarray}
where $\varphi(x)$ and $\Pi(x)$ are canonically conjugate bosonic fields and $K$ is related to the coupling $g$ in a non-universal way. It will be necessary later to compare the bosonic and fermionic approaches which requires us to determine this relation between $K$ and $g$ or more precisely, between $K$ and $\phi$. To do so we compute the compressibility in the fermionic language and match it to the known result from bosonization. With this in mind we note that the linear spectrum of the fermionic model \eqref{H} means we need to impose a momentum cutoff of $-\mathcal{D}$ and construct the ground state by populating states from this level up. Therefore in the thermodynamic limit the ground state energy for density $D=N/L$ with a chemical potential $\mu$ is,
\begin{eqnarray}\label{ell}
\frac{E}{L}=\int_{-\mathcal{D}}^{2\pi D-\mathcal{D}}\left[ k-\left(\phi\int_{-\mathcal{D}}^{2\pi D-\mathcal{D}} \frac{\mathrm{d}q}{2\pi}\right)-\mu\right]\frac{\mathrm{d}k}{2\pi}.
\end{eqnarray}
Varying the density with  both the cutoff and $\mu$ held fixed we  can find the compressibility of the bulk. Sending  $D\to D+\delta D$ and  minimizing $E$ with respect to $\delta D$ we get $2\pi(1-\frac{\phi}{\pi})D=\mathcal{D}+\mu$ and so the  compressibility is related to $\phi$ by
$\phi/\pi=1-\varkappa/\varkappa_0$
where we denote the free compressibility by $\varkappa_0=1/2\pi$ and that of the Luttinger liquid $\varkappa$. In turn this can be expressed in terms of the Luttinger liquid parameters 
\begin{eqnarray}\label{comp}
\frac{\phi}{\pi}=1-\frac{1}{K}.
\end{eqnarray}
 We replace the renormalization scheme dependent  coupling 
 $g$  with  the scattering phase    $\phi$  which can be directly related to measurable quantities via     \eqref{comp}. Note that being a phase $\phi$ is  restricted to lie in the interval $[-\pi, \pi]$ and therefore the fermionic Hamiltonian \eqref{H} can only realize $K\in [1/2,\infty]$.

The Luttinger wire is attached to a quantum dot  modeled by a resonant level with energy $\epsilon_0$   via a tunneling term $t$ \cite{FurMat}. They are further coupled via a Coulomb interaction $U$,
\begin{eqnarray}\label{Ht}
H_t&=& \frac{t}{2} (\psi^\dagger_+(0)-\psi^\dagger_-(0)) d +\text{h.c},\\\label{HU}
H_{d}&=&\epsilon_0 d^\dag d+\frac{U}{2}d^\dag d\sum_{\sigma=\pm}\psi^\dag_\sigma(0)\psi_\sigma(0).
\end{eqnarray}
When coupled to the dot  the conservation law takes the form  $\hat{N}=\hat{N}_++\hat{N}_-+\hat{n}_d$,  the total particle number (here $\hat{n}_d= d^{\dagger}d$).  

\section{The Eigenstates}

We will proceed with the diagonalization of $H=H_{LL}+H_t+H_d$ in the usual Bethe Ansatz manner by first finding the single particle eigenstates, then the two particle states from which we deduce the $N$ particle solution. Following this the spectrum is determined in terms of the Bethe Ansatz equations by imposing boundary conditions on the system.
 
We again unfold the system as before and write the most general single particle state of energy $k$ as 
\begin{eqnarray}
\ket{k}=\int e^{ikx}\left[A^{[10]} \theta(-x)+A^{[01]} \theta(x)\right]\psi^\dag(x)\ket{0}+Bd^\dag\ket{0}
\end{eqnarray}
Upon acting on this state with the Hamiltonian we find it is an eigenstate provided,
\begin{eqnarray}\label{1phase}
S^{10}=\frac{A^{[01]}}{A^{[10]}}&=&\frac{k-\epsilon_0-i\Gamma}{k-\epsilon_0+i\Gamma},\\
B&=&\frac{t}{k-\epsilon_0}\left(A^{[10]}+A^{[01]}\right).
\end{eqnarray}
The quantity $\Gamma=t^2/2$ is the hybridization width  while $S^{10}$ is the single particle S-matrix  for fermion  scattering past the dot. Moving to the two particle case the interaction parameters $U$ and $g$ enter into play. We can write  the state with energy $E=k_1+k_2$ as 
\begin{eqnarray}\nonumber
\ket{k_1,k_2}=\int \sum_Q A^Q\theta(\vec{x}_Q)e^{ik_{1}x_1+ik_2x_2}\psi^\dag(x_1)\psi^\dag(x_2)\ket{0}\\
+\int \sum_P\left[B_1^Pe^{ik_1x}+B_2^Pe^{ik_2x}\right]\theta(x_P)\psi^\dag(x)d^\dag\ket{0}.\quad
\end{eqnarray} 
Here, in the first line we have expanded the two fermion part of the wavefunction into 8 regions which contain every ordering of the particles in addition to distinguishing whichever is closest to the origin, labelled by $Q\in\{[120],[210],[012],[021],[102A],[102B],[201A],[201B]\}$.  For example the amplitude $A^{[102B]}$ corresponds to the region with $x_1<0<x_2$ and $|x_1|>|x_2|$ whereas $A^{[102A]}$ has $|x_2|>|x_1|$. The $\theta(\vec{x}_Q)$ are Heaviside functions which are non zero only in the region $Q$. These extra regions compared to standard Bethe wavefunctions are required by the non local interaction and are necessary when studying Luttinger impurity models \cite{ryl}. In the second line, the wavefunction in the dot part is expanded in regions $P$ which correspond to the fermion being either to the left or to the right of the origin e.g. $B^{[10]}_2$ is the amplitude for the particle with $k_2$ to the left of the dot while the other particle is on it.

 Acting on this state with the Hamiltonian we find it is an eigenstate provided,
\begin{eqnarray}
\frac{A^{[201A]}}{A^{[210]}}&=&\frac{A^{[012]}}{A^{[102A]}}=\frac{k_1-\epsilon_0-i\Gamma}{k_1-\epsilon_0+i\Gamma},\\
\frac{A^{[102B]}}{A^{[120]}}&=&\frac{A^{[021]}}{A^{[201B]}}=\frac{k_2-\epsilon_0-i\Gamma}{k_2-\epsilon_0+i\Gamma},\\\label{lphase}
\frac{A^{[102A]}}{A^{[102B]}}&=&\frac{A^{[201B]}}{A^{[201A]}}= e^{i\phi}  \\
\frac{A^{[210]}}{A^{[120]}}&=&\frac{A^{[021]}}{A^{[012]}}=  S^{12}
\end{eqnarray}
with
\begin{eqnarray}\label{2phase}
S^{12}=\frac{k_1+k_2-2\bar{\epsilon}_0-i\frac{U'}{2}\left(k_1-k_2\right)}{k_1+k_2-2\bar{\epsilon}_0+i\frac{U'}{2}\left(k_1-k_2\right)}
\end{eqnarray}
being  the S-matrix  when a particle of energy $k_1$ scatters past one of energy $k_2$, and we defined,
\begin{eqnarray}
\arctan{(U'/2)}&=&\arctan{(U/2)}-\arctan{(g)}\nonumber \\
\bar{\epsilon}_0&=&\epsilon_0-\Gamma U'/2.\nonumber
\end{eqnarray}
The parameters $U'$ and $\bar{\epsilon}_0$ are bare quantities  and as such depend upon the regularisation scheme employed. These parameters must be related to universal quantities to acquire meaning as is always the case for renormalisable field theories. Below we relate $U'$ to $K$ and $\bar{\epsilon}_0$ to the renormalised dot energy.

Generalising to $N$ particles, the state consists of parts with the dot occupied or unoccupied. The latter is written as
\begin{eqnarray}\label{N}
\ket{\vec{k}}=\sum_Q\int A^Q\theta(\vec{x}_Q) e^{\sum_j^Nk_jx_j}\prod_{j=1}^N\psi^{\dag}(x_j)\ket{0}.
\end{eqnarray}
The sum is now over $2^NN!$ regions $Q$ and the amplitudes are related to each by the various phase shifts given in \eqref{1phase}, \eqref{2phase} and \eqref{lphase}. The occupied dot part can also be written in such a fashion, we omit it here as we will only require  \eqref{N} to proceed. The consistency of the solution is guaranteed as the S-matrices satisfy the reflection equation \cite{Sklyannin},
\begin{eqnarray}
S^{k0}e^{i\phi}S^{j0}S^{jk}=S^{jk}S^{j0}e^{i\phi}S^{k0}
\end{eqnarray}
along with the Yang Baxter equation $S^{ki}S^{ji}S^{jk}=S^{jk}S^{ji}S^{ki}$, which they do so trivially as all the S-matrices are phases. 

The  $k$ dependent two body S-matrix \eqref{2phase} is the same form as the interacting resonant level model (IRLM) which describes a dot coupled to Fermi liquid leads \cite{IRL2}. The effect of the bulk interaction on this is  to shift $U\to U'$. This makes explicit the relationship between the IRLM and the Luttinger resonant level model seen in \cite{GWB}, that is, when only the thermodynamics of the dot are concerned one can deal with the level-lead interaction instead of a bulk interaction. Bulk properties, however, differ in both models as do the structure of the wave functions which will show up as different correlation functions.

To determine the spectrum we impose periodic boundary conditions in the unfolded system  which  as stated before corresponds to an open boundary condition at $x=-L/2$ in the folded language. Upon doing so we find the Bethe equations which determine the $k_j$,
\begin{eqnarray}\nonumber
e^{-ik_jL }&=&e^{i(N-1)\phi}\frac{k_j-\epsilon_0-i\Gamma}{k_j-\epsilon_0+i\Gamma}\\\label{BAE}
&\times&\prod_{l}^N\frac{k_j+k_l-2\bar{\epsilon}_0-i\frac{U'}{2}\left(k_j-k_l\right)}{k_j+k_l-2\bar{\epsilon}_0+i\frac{U'}{2}\left(k_j-k_l\right)}.
\end{eqnarray}
The interpretation of these in the folded system is an incoming  right mover incident from the left, moving toward the dot and scattering past the other particles in the system.  When the other particle is an outgoing left mover a constant phase $e^{i\phi}$ is acquired whereas if it goes past another incoming particle it gains the $k$ dependent two particle phase shift \eqref{2phase}.  After scattering off the dot and picking a factor as in \eqref{1phase}, the particle moves back across the system as a left mover this time picking up $e^{i\phi}$ from the remaining incoming particles and  \eqref{2phase}  from the other outgoing left movers.  

We conclude this section by remarking that the coupling of the dot to the bulk system has caused two differences in the Bethe equations as compared to the $e^{-ik_jL}=e^{i(N-1)\phi}$ we found above for a Luttinger liquid in a box. These are the inclusion of the dot phase shift and the $k$ dependent two particle phase shift. As we will see below the latter plays a crucial role in determining the behavior of the system impacting the thermodynamic properties and its RG flow.

Moreover we would like to comment that the relation between $K$ and $\phi$ obtained before is still valid despite the inclusion of this new two particle phase shift. To see this we drop the dot term in the Bethe equations and take their log to recover the Luttinger liquid energy,
\begin{eqnarray}
E=\frac{2\pi}{L}\sum_j^N n_j-N(N-1)\frac{\phi}{L}
\end{eqnarray}
with $n_j$ being integers. The log of the two particle phase shift is odd and therefore cancels out when summed over all particles. This is the discrete form of \eqref{ell} and we could proceed as we did before to obtain the same relation. 

\section{Zero Temperature properties}

Having obtained the Bethe equations, \eqref{BAE}, we seek to identify  the ground state of the system. This is most easily accomplished by describing the particles in terms of their rapidity $x_j$ defined by $k_j=\mathcal{D}e^{x_j}+\bar{\epsilon}_0$, where $-\mathcal{D}$ is the lower momentum cutoff. The energy is now: $E=\sum_j^N    \mathcal{D}e^{x_j}+N\bar{\epsilon}_0$ and \eqref{BAE} becomes,
\begin{eqnarray}\nonumber
e^{-i\mathcal{D}e^{x_j}L}\! \!&=&\!\!e^{i(N-1)\phi+i\epsilon_0L}\frac{\cosh{\frac{1}{2}\left(x_j-c+i\Delta\right)}}{\cosh{\frac{1}{2}\left(x_j-c-i\Delta\right)}}\\\label{Bae}
&&\times\prod_{l}^N\frac{\sinh{\frac{1}{2}\left(x_j-x_l-2i\Delta\right)}}{\sinh{\frac{1}{2}\left(x_j-x_l+2i\Delta\right)}}
\end{eqnarray}
The parameters $\Delta$, $c$ and $\phi$ encode the interactions in the model and the effect of the dot, they are defined as  
\begin{eqnarray}\label{c}
e^c&=&\gamma\frac{\Gamma}{\mathcal{D}}\\
 \Delta &=&\frac{\pi}{2}\left(2-\frac{1}{K}\right)+\arctan{(\frac{U}{2})}.
 \end{eqnarray}
with $\gamma=1/\sqrt{1+(U'/2)^2}$. Here we see that the presence of the $U$ contributes to a local modification of $\phi$ or $K$ in the bosonised language. This  could be understood physically by integrating out the dot degrees of freedom, whereupon the interaction term in the Hamiltonian is modified locally near the dot. Bulk properties are still dependent only on $\phi$ or $K$  but dot quantities like the occupation calculated below depend on $\Delta$. At both $\Delta=0$ and $\Delta=\pi/2$ the two particle phase shift vanishes and the system simplifies considerably. If we set $U=0$, the latter corresponds to the resonant level model of free fermions coupled to the dot. At $\Delta=0$ however the single particle phase shift also vanishes, this value corresponds to maximally repulsive fermions (again for $U=0$) which causes the system to seize and prevent any tunneling to the dot. Aside from these two free points we will see below that at $\Delta=\pi/3$  the nature of the ground state changes. 
\begin{figure}
\centering
\includegraphics[trim = 0mm 55mm 0mm 0mm, clip, width=0.49\textwidth]{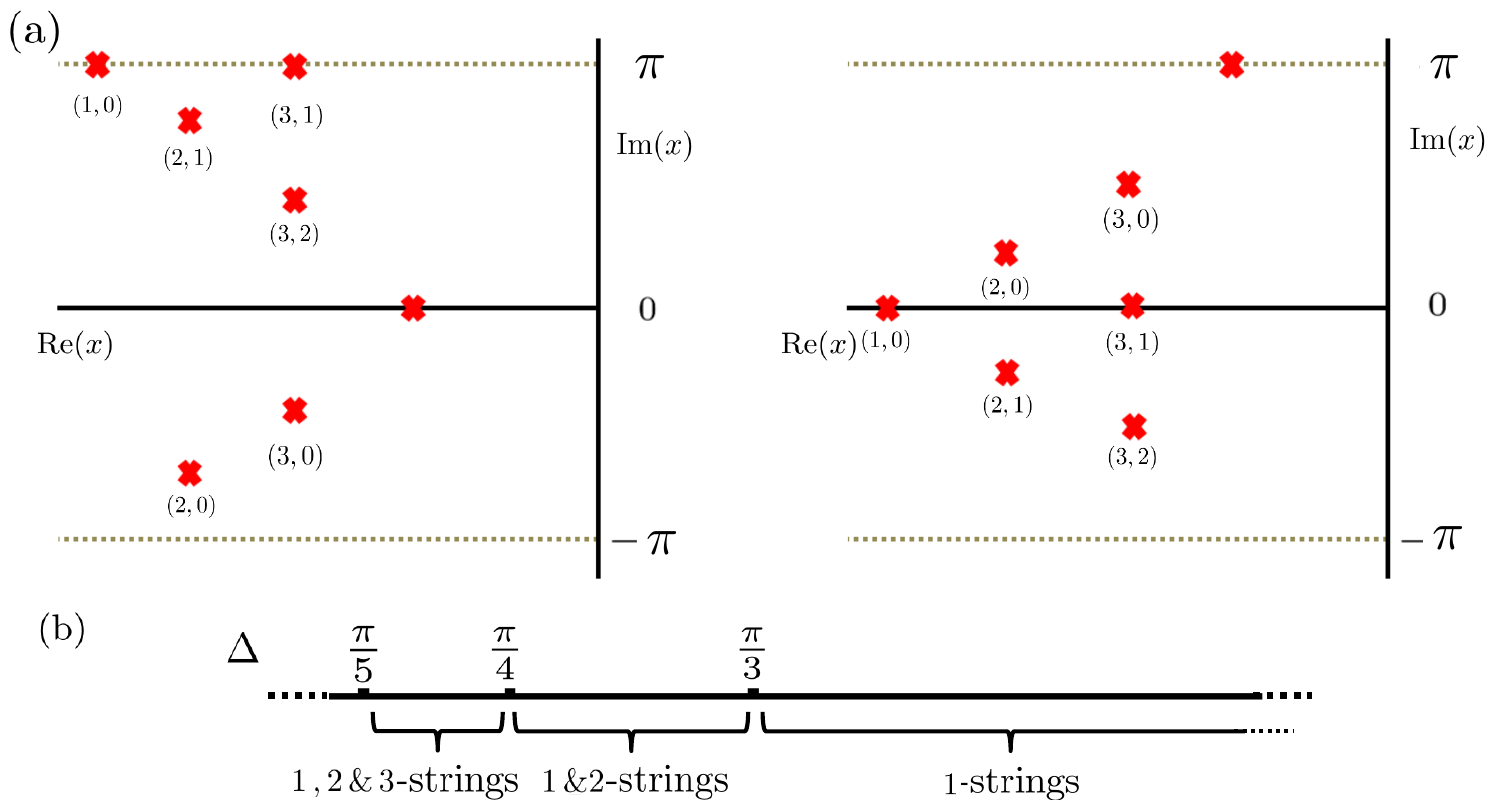}
\caption{(a) The configurations of allowed strings for $\pi/4\leq\Delta<\pi/3$ on the left and  for $2\pi/3<\Delta\leq 3\pi/4$ on the right. In both cases strings of length up  $n\leq 3$ are allowed, as well as  additional 1-strings corresponding to a positive/negative energy particle. Red crosses mark the string elements and underneath each  $(n,l)$ denotes the string length and the element of the string (see text). On the left,  the spacing between adjacent elements of a string, i.e  between $l$ and $l+1$  for fixed $n$,  is $i\Delta$  and the elements are  symmetrically placed (modulo $2\pi$) with respect to $i \pi$ axis, in addition to real 1-strings.  For the  strings on the right, the spacing is  $i(\pi-\Delta)$, the elements are  symmetrically placed around the real axis and there are 1-strings occupying the $i \pi$ axis. 
 (b) The form of the ground state depends on the regime in which $\Delta$ lies. For $\Delta>\pi/3$ it consists of $1-$ strings only,  below this it changes to consisting of $1-$ and $2-$ strings and then to include $3-$strings and so on.}
\end{figure}
\subsection{Identifying the ground state}
To identify the ground state of the system we must list the  possible types of solutions to the Bethe equations \eqref{Bae}. In order to do so we note that apart from the dot term the Bethe equations are similar to those of the massive Thirring model which have been widely studied \cite{BT}\cite{Korepin1}\cite{Korepin2}\cite{Fowler} and in fact can be thought of as a massless limit of these \cite{FS}\cite{FSW}. This massless limit is known not to change the possible types of solutions known as strings which depend upon $\Delta$ and we now list. First consider $\Delta>\pi/2$ and in particular take
\begin{equation}
\pi\frac{\nu-1}{\nu}< \Delta\le\pi\frac{\nu}{\nu+1}
\end{equation} 
with $\nu\ge 2$ a positive integer. In this region the rapidities can be complex and form so called $n$-strings such that $x^l=x+i(\pi-\Delta)(n-1-2l)$ with $x$ real, $l=0,\dots,n-1$ and $0\leq n\leq \nu$. These $n$-strings can be thought of as bound states and have positive bare energy $E_n=\sin{(n(\pi-\Delta))}\mathcal{D}e^x/\sin{(\Delta)}$. Additionally there are negative energy particles that have Im$(x)=\pi$. For $\Delta \le \pi/2$  the range slips into regions,
\begin{equation}\label{Dl}
\frac{\pi}{\nu+1}\le\Delta<\frac{\pi}{\nu}
\end{equation} 
in which the $n$-strings  take the different form $x^l=x+i\pi+i\Delta(n-1-2l)$, $l=0,\dots,n-1$ and $n\leq \nu$. The $n$-strings now have negative bare energy  $E_n=-\sin{(n\Delta)}\mathcal{D}e^x/\sin{(\Delta)}$ and are in addition to positive energy particles which have real rapidity. The arrangement of the allowed strings for two values of $\Delta$ are shown explicitly in Figure 2 (a) \footnote{Deviations to these string solutions as well solutions which fall outside this class are known to exist and are important when studying the completeness of the Bethe ansatz eigenstates as well as when correlation functions are considered \cite{Essler}\cite{Isler}\cite{Kundu}\cite{Deguchi}\cite{Hagemans}\cite{Destri}. For our purposes however we shall simply use the strings as presented above. }. 

We now proceed to construct the ground state following \cite{Korepin1} which  consists of all possible negative energy particles filled from the cutoff,$-\mathcal{D}$, upwards.  We begin by considering the regime  $\Delta\geq\pi/3$ where only one type of negative energy particle is available (below $\pi/2$  2-strings are also allowed but these can be shown to increase the energy). Therefore we set Im$(x_j)=\pi$ in \eqref{Bae}
and take the thermodynamic limit by sending $N,L\to \infty$   while the  cutoff $\mathcal{D}$  is held fixed at a value larger  than all  quantities such as   $\bar{\epsilon}_0, \Gamma$. The density, $D=N/L$ is then obtained by minimizing the energy for a given large $\mathcal{D}$. In this limit the particle rapidities $x_j$ approach each other and can be described by a continuous distribution $\rho^p(x)$,  the Bethe density of states. Similarly the distribution of holes is denoted $\rho^h(x)$. The Bethe equations become an integral equation determining  these distributions,

\begin{eqnarray}\nonumber
   \frac{1}{2\pi}\mathcal{D}e^x+\frac{1}{L}a_1(x-c)=\rho^p(x)+\rho^h(x) \quad\quad\quad
\\   \label{bethgs}
   \quad\quad\quad+\int_{-B}^0a_2(x-y)\rho^p(y)\\
a_n(x)=\frac{i}{2\pi}\d{}{x}\log{\frac{\sinh{\frac{1}{2}(x+in\Delta})}{\sinh{\frac{1}{2}(x-in\Delta)}}},
\end{eqnarray} 
where the lower integration limit $B$ depends on $\bar{\epsilon}_0$ and is determined by minimizing the energy with the dot energy fixed. This then determines the hole distribution $\rho^h(x)$.

If we set $\bar{\epsilon}_0=0$ then no holes appear in the ground state meaning $\rho^h(x)=0$ and $B=\infty$. Since we are interested in the physics at scales well below the cutoff $\mathcal{D}$ which we later send to $\infty$, we need only be concerned with rapidities $x\ll 0$. The 
ground state distribution,  denoted $\rho^0(x)$ can therefore be found by Fourier transform giving
\begin{eqnarray}
\label{gs 0}
\rho^0(x)=\frac{\tan{(\frac{\pi^2}{2\Delta})}}{\pi-2\Delta}\frac{\mathcal{D}}{2\pi}e^{\frac{\pi}{2\Delta}x}+\frac{1}{L}s(x-c)
\end{eqnarray}
with $s(x)=1/(4\Delta\cosh(\pi x/2\Delta))$. The first term of $\rho^0(x)$ is the bulk contribution and the second is due to the dot.

To confirm that that this is indeed the  ground state of the system for $\Delta\geq \pi/3$ we can construct excitations and check that they increase the energy. The simplest type of excitation consists of adding a hole to the ground state. As with many other Bethe Ansatz models, the energy: $\varepsilon^h(x)$,  of this excitation is proportional the ground state distribution,
\begin{eqnarray}
\varepsilon^h(x)=2\pi\rho^0(x)>0.
\end{eqnarray}
Other excitations consist of adding $n$-strings or positive energy particles which can also be shown to increase the energy.

We now consider the parameter regime, $\Delta<\pi/3$.  The availability  of additional negative energy particles in this regime changes the nature of the ground state \cite{Korepin2}. More  specifically, for values of $\Delta$
specified by  \eqref{Dl} the ground state consists of all $n$-strings  for $n\leq \nu-1$ filled from the cutoff upwards;  e.g for $\pi/4\le\Delta<\pi/3$ the ground state consists of both 1- and 2-strings, while for $\pi/5\le\Delta<\pi/4$ the ground state consists of all possible 1-, 2- and 3-strings, see FIG. 4(b).  Inserting these configurations into \eqref{Bae} and taking the thermodynamic limit  the Bethe equations become $\nu-1$ coupled integral equations for the $n$-string particle and hole distributions  $\rho_j^p(x),\rho_j^h(x)$,
\begin{eqnarray}\nonumber
\frac{\sin{(n\Delta)}}{\sin{(\Delta)}}\frac{\mathcal{D}}{2\pi}e^x+\frac{1}{L}a_n(x-c)=\rho^p_n(x)+\rho^h_n(x)\quad\quad\quad\\\label{Bael}+\sum_{k}^{\nu-1}\int_{-B}^0\mathbb{T}_{nk}(x-y)\rho_k(y)\quad\quad\quad
\end{eqnarray}
Here $\mathbb{T}_{nk}=a_{n+k}(x)+a_{k-n}(x)+2\sum_{l=1}^{n-1}a_{k-n+2l}(x)$ is the derivative of the phase shift between strings of length $n$ and $k$ with $n<k$ and has the property $\mathbb{T}_{j,k}=\mathbb{T}_{k,j}$. Also, as before $B$ must be determined by minimizing the energy with $\bar{\epsilon}_0$ held fixed.

 We first analyse the system with $\bar{\epsilon}_0=0$ where again there are no holes in the ground state and $B=\infty$. The solution is obtained by inverting the matrix $1+\mathbb{T}$ \cite{Korepin2},
\begin{eqnarray}\nonumber
(1+\mathbb{T})_{jk}^{-1}=\delta_{jk}(\delta(x)-\delta_{k,\nu-1}b(x))-\left(\delta_{j,k+1}\right.
\\\label{Tinv}
\left.\,\,+\delta_{j,k-1}\right)s(x)\\
\tilde{b}(\omega)=\frac{\sinh{[(\pi-\nu\Delta)\omega]}}{2\cosh{(\Delta\omega)}\sinh{[\pi-(\nu-1)\Delta)\omega]}}.
\end{eqnarray}

Applying this to \eqref{Bael} we obtain the ground state distributions, 
\begin{eqnarray}
\rho^0_n(x)=d_n\frac{\mathcal{D}}{2\pi}e^{\frac{\pi}{2\Delta}x}+\delta_{j,1}\frac{1}{L}s(x-c)
\end{eqnarray}
where the coefficients $d_n$ are 
\begin{eqnarray}\label{gs l}
d_{n}&=&\frac{1}{\pi-2\Delta}\left(\frac{2\sin{(n\Delta)}}{\tan{(\Delta)}}\right)~~~\text{for}~n<\nu-1\\\nonumber
d_{\nu-1}&=&\frac{1}{\pi-2\Delta}\left(\frac{\sin{((\nu-2)\Delta)}}{\sin{(\Delta)}}\right.\\
&&\left.+\frac{\sin{((\nu-1)\Delta)}}{\sin{(\Delta)}}\tan{(\pi-\nu\Delta)\frac{\pi}{2\Delta}}\right)
\end{eqnarray}
Note that the dot contribution  appears only  in the distribution of $1$-stings $\rho^0_1(x)$ and is the same as for $\Delta\geq\pi/3$ \eqref{gs 0}.  Again, to verify this is the ground state  we show that any modification results  in excitations that increase the energy. The simplest type of excitation is adding a hole to the $n$-string distribution. Just as before the energy of this is given by
\begin{eqnarray}
\varepsilon^h_n(x)=2\pi\rho^0_n(x)>0.
\end{eqnarray} 
Other excitations consist of adding $\nu$-strings or positive energy particles which can be also checked to increase the energy.

\subsection{The Dot Occupation}

In this section we calculate the ground state occupation of the dot  $n_d=\left<d^\dag d\right>$ as a function of the dot energy $\bar{\epsilon}_0$ and $\Delta$. The non zero dot energy means that the ground state will contain holes as well as particles and furthermore that $B$ is finite. To determine $B$ we recall that the energy is given generically by $E=-\mathcal{D}\sum_je^{x_j}+N\bar{\epsilon}_0$ and that the ground state is found by balancing the energy cost due to the second term with that of a hole. Therefore given that $\varepsilon^h_n(x),\varepsilon^h(x)\propto \mathcal{D}e^{\frac{\pi}{2\Delta}x}$ we have
\begin{eqnarray}\label{B}
\bar{\epsilon}_0=\alpha \mathcal{D}e^{-\frac{\pi}{2\Delta}B}
\end{eqnarray}
where $\alpha$ is a positive constant whose value depends on the regime $\Delta$ lies in (see appendix for more details).  
\begin{figure}
\centering
\includegraphics[trim = 0mm 41mm 0mm 0mm, clip, width=0.48\textwidth]{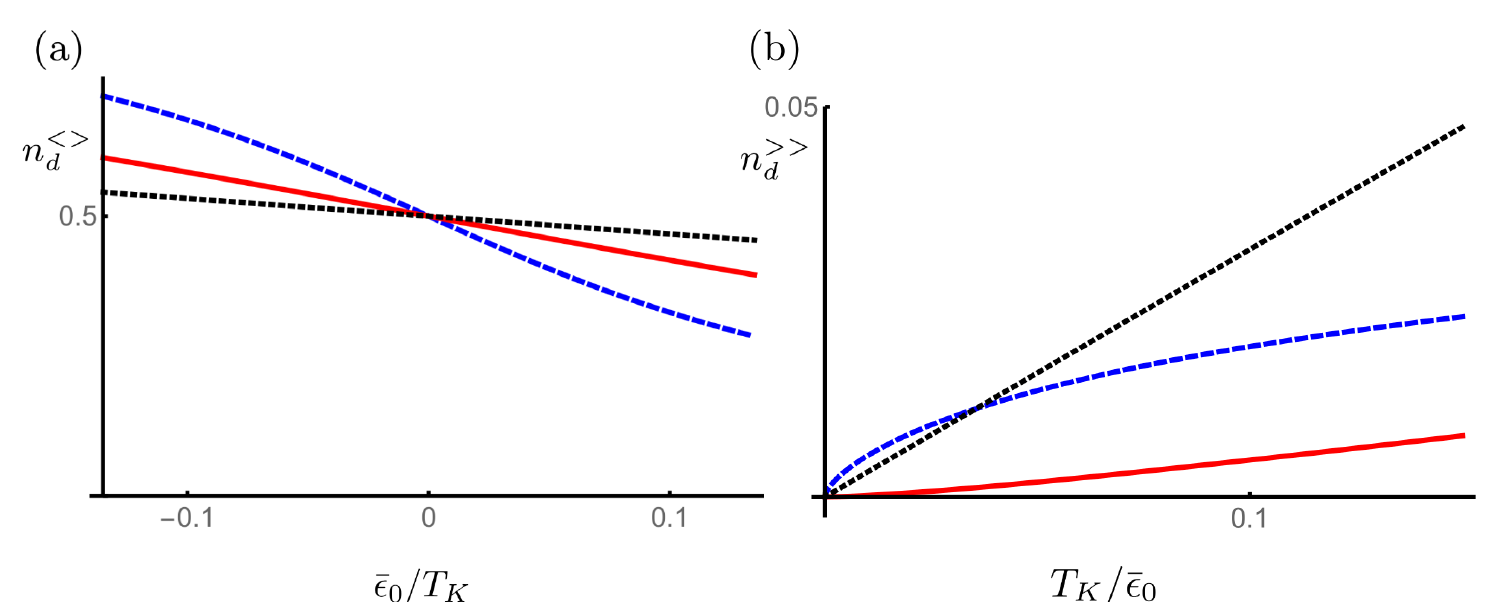}
\caption{(a) The dot occupation, $n^{<>}_d$,  as a function of $\bar{\epsilon}_0/T_K$ for $\Delta=\pi/3$ (dashed, blue), $\Delta=\pi/2$ (dotted black) and $\Delta=3\pi/4$ (solid, red) from \eqref{ndgl} (b) $n_d^{>>}$ from \eqref{ndgg} as function of $T_K/\bar{\epsilon}_0$  for $\Delta=\pi/3$ (dashed, blue), $\Delta=\pi/2$ (dotted black) and $\Delta=3\pi/4$ (solid, red) from \eqref{ndgg}. Recall that for $\Delta=\pi/2$ the system interactions simplify considerably,  corresponding to $K=1/2$ (maximally repulsive) for $U=0$.}
\end{figure}

Since the ground state differs considerably above and below $\Delta=\pi/3$ we will employ two different methods to find $n_d$. We begin with the region $\Delta\geq \pi/3$ and obtain the desired quantity by integrating over the dot contribution to the density of states, 
\begin{eqnarray}
n_d&=&\int_{-B}^0\rho^p_d(x)\mathrm{d}x\\\label{baedg}
a_1(x-c)&=&\rho^p_d(x)+\int_{-B}^\infty a_2(x-y)\rho^p_d(y)\mathrm{d}y
\end{eqnarray}
The second line is obtained by extracting the dot dependent quantities from \eqref{bethgs} and extending the upper integral limit to $\infty$ which can be done as the driving term is localised about $x=c\ll 0$. The dot distribution can be found by means of the Wiener-Hopf method (See \cite{TWAKM}, \cite{RMP}  or \cite{Takahashi} and references therein). We factorize the kernel into factors $G_{\pm}(\omega)$ that are analytic in the upper and lower half planes,  $G_+(\omega)=G_-(-\omega)$,
\begin{eqnarray}\label{G}
G_+(\omega)&=&\frac{\Gamma(\frac{1}{2}-i\frac{\Delta}{\pi}\omega)\Gamma(1-i\frac{\pi-\Delta}{\pi}\omega)  }{\sqrt{2(\pi-\Delta)}\Gamma(1-i\omega)}e^{i\omega a},\\
a&=&\left(\frac{\pi-\Delta}{\pi}\right) \log\left(\frac{\pi-\Delta}{\Delta}\right)-\log{\left(\frac{\pi}{\Delta}\right)}.
\end{eqnarray}
where $\Gamma(x)$ is the Gamma function.  Then, noting that $n_d=\tilde{\rho}^p_d(0)$ we find
\begin{eqnarray}\label{n}
n_d=\frac{-i}{2\pi }G_+(0)\int_{-\infty}^{\infty}\frac{G_-(\omega)\tilde{a}_1(\omega)}{\omega-i0}e^{i\omega(c+B)}.
\end{eqnarray}  
which can be evaluated by closing the contour in the upper or lower half plane depending upon the sign of $c+B$. Having determined $B$ through \eqref{B} we have that, 
\begin{eqnarray}
c+B&=&\frac{2\Delta}{\pi}\log\left(\frac{T_K}{\bar{\epsilon}_{0}}\right)\\\label{Tk}
T_K&\equiv& \alpha \mathcal{D}\left(\gamma\frac{\Gamma}{\mathcal{D}}\right)^{\frac{\pi}{2\Delta}}
\end{eqnarray}
where we have defined the strong coupling scale $T_K$. All physical energies are measured with respect to this scale which has been dynamically generated by the model. We hold it fixed while taking $\mathcal{D}\to \infty$ thereby obtaining universal results. The form of $T_K$ will be discussed further below. 

We now proceed to obtain  expressions for the dot occupation using \eqref{n}. By closing the contour in the upper half plane we determine the expansion for $\bar{\epsilon}_0<T_K$  (and $\Delta\ge \pi/3$) which we denote $n^{<>}_d$,
\begin{eqnarray}\nonumber
n_d^{<>}(\bar{\epsilon}_0,\Delta)=\frac{1}{2}-\frac{1}{\sqrt{\pi}}\sum_{n=0}^\infty\frac{(-1)^n}{n!}\frac{e^{\frac{\pi}{2\Delta}(2n+1)a}}{2n+1}\left(\frac{\bar{\epsilon}_0}{T_K}\right)^{2n+1}\\\label{ndgl}
\times\frac{\Gamma(1+\frac{\pi}{2\Delta}(2n+1))}{\Gamma(1+\frac{\pi-\Delta}{2\Delta}(2n+1))}.\quad\quad
\end{eqnarray}
 On the other hand, closing the contour in the lower half plane we get the occupation when the dot energy is larger than the strong coupling scale, $\bar{\epsilon}_0 \ge T_K$. Denoting this $n_d^{>>}(\bar{\epsilon}_0,\Delta)$,   the expansion is now,
\begin{eqnarray}\nonumber
n_d^{>>}(\bar{\epsilon}_0,\Delta)=\frac{1}{2\sqrt{\pi}}\sum_{n=1}^\infty\frac{(-1)^{n+1}}{n!}e^{-n a}\frac{\Gamma(\frac{1}{2}+\frac{\Delta}{\pi}n)}{\Gamma(1-\frac{\pi-\Delta}{\pi}n)}\\\label{ndgg}
\times\left(\frac{T_K}{\bar{\epsilon}_0}\right)^{\frac{2\Delta}{\pi}n}.
\end{eqnarray}
 The dot occupation is plotted for some values of $\Delta$ in FIG. 3 where we have used $n_d(-\bar{\epsilon}_0)=1-n_d(\bar{\epsilon}_0)$ \cite{FurMat} to obtain the expressions for negative dot energy.
 
To find the expressions analogous to \eqref{ndgl} and \eqref{ndgg} in the region $\Delta<\pi/3$ we employ a different method. Starting from \eqref{Bae} it can be shown that the dot contribution to ground state energy of the system is,
\begin{eqnarray}\nonumber
E_d
=-\int^{-B}_{-\infty}S(x-c)\rho_1^h(x)
\end{eqnarray}
where $S'(x)=s(x)$. The dot occupation is therefore given by,
\begin{eqnarray}
n_d=\frac{1}{2\pi}\pd{}{\bar{\epsilon}_0}\int_{-\infty}^\infty\frac{\tilde{s}(\omega)}{i\omega}\tilde{r}_1(\omega)e^{-i\omega(c+B)}\mathrm{d}\omega
\end{eqnarray}
where we have defined $r_n(x)=\rho^h_n(x-B)$ with $B(\bar{\epsilon}_0)$  already determined.  Now to evaluate this explicitly one needs to solve \eqref{Bael} for the hole distributions which cannot be achieved analytically. We can however determine the positions of its zeros and poles. Given that $\rho^h_1(x)=0$ for $x>-B$ we know that  $\tilde{r}_1(\omega)$ is analytic in the lower half plane and additionally $r_1(x)\propto \mathcal{D}e^{-\frac{\pi}{2\Delta}B}$. Furthermore the zeros and poles of $\tilde{r}_1^h(\omega)$ are fixed by the poles and zeros of the determinant of $1+\mathbb{T}$ respectively \cite{Korepin2}. Thus it has zeros at $i\pi(n+1/2)/\Delta$ and poles at $i(n+1)$. Combining all this we find  the dot occupation for $\Delta<\pi/3$. For small $\bar{\epsilon}_0<T_K$ we denote it $n_d^{<<}$,
\begin{eqnarray}\nonumber
n_d^{<<}(\bar{\epsilon}_0,\Delta)=\frac{1}{2}+\pd{}{\bar{\epsilon}_0}\sum_{n=0}^\infty(-1)^n \frac{\tilde{r}_1(-i\frac{\pi}{2\Delta}(2n+1))}{\pi(2n+1)}\\\label{ndll}
\times\left(\frac{\bar{\epsilon}_0}{T_K}\right)^{2n+1}
\end{eqnarray}
At large dot energy $\bar{\epsilon}_0>T_K$, the dot occupation in the    $\Delta<\pi/3$ regime,  $n_d^{><}$ ,  is obtained by closing the contour in the upper half plane. We find,
\begin{eqnarray}\label{ndlg}
n_d^{><}(\bar{\epsilon}_0,\Delta)=\sum_{n=0}^\infty a_n\left(\frac{T_K}{\bar{\epsilon}_0}\right)^{\frac{2\Delta}{\pi}(n+1)}
\end{eqnarray}
for some constants $a_n$ which depend on the residues of $\tilde{r}_1(\omega)$.

\subsection{The RG Flow}

In the preceding sections we have derived the dot occupation in the ground state as a function of $\Delta$ and  $\bar{\epsilon}_0/T_K$ with $T_K$ being a strong coupling scale generated by the model. 

The dynamic generation of a scale  $T_K$, akin to the Kondo scale, can be understood in this spinless model by making the analogy between  the charge fluctuations on the dot and the spin fluctuations in the Kondo model. By identifying the impurity spin and dot occupation via $S^z=n_d-1/2$, a screened Kondo spin corresponds to a fully hybridized dot with fixed occupation, $n_d=1/2$ while the unscreened spin corresponds to the dot being decoupled and therefore being either full or empty, $n_d=0,1$. The role of an external magnetic field in the Kondo model is fulfilled here by the dot energy $\bar{\epsilon}_0$.  We will now discuss appearance of these regimes in our model.

 In order to obtain universal results we have held $T_K$  fixed while removing the cutoff $\mathcal{D}\to\infty$ having previously assumed all scales are much smaller than $\mathcal{D}$. In particular we must have $T_K\ll \mathcal{D}$ and so to fulfil this we need $\Delta>0$.  For $\Delta<0$ on the other hand there is no universal regime as the would-be scale is above the cutoff and  universal results cannot be obtained. If we set $U=0$ then this transition between universal and non-universal regimes occurs at $K=1/2$ and is shifted by a non zero $U$ in agreement with perturbation theory \cite{FurMat}.

\begin{figure}
\centering
\includegraphics[trim = 0mm 20mm 0mm 0mm, clip, width=0.35\textwidth]{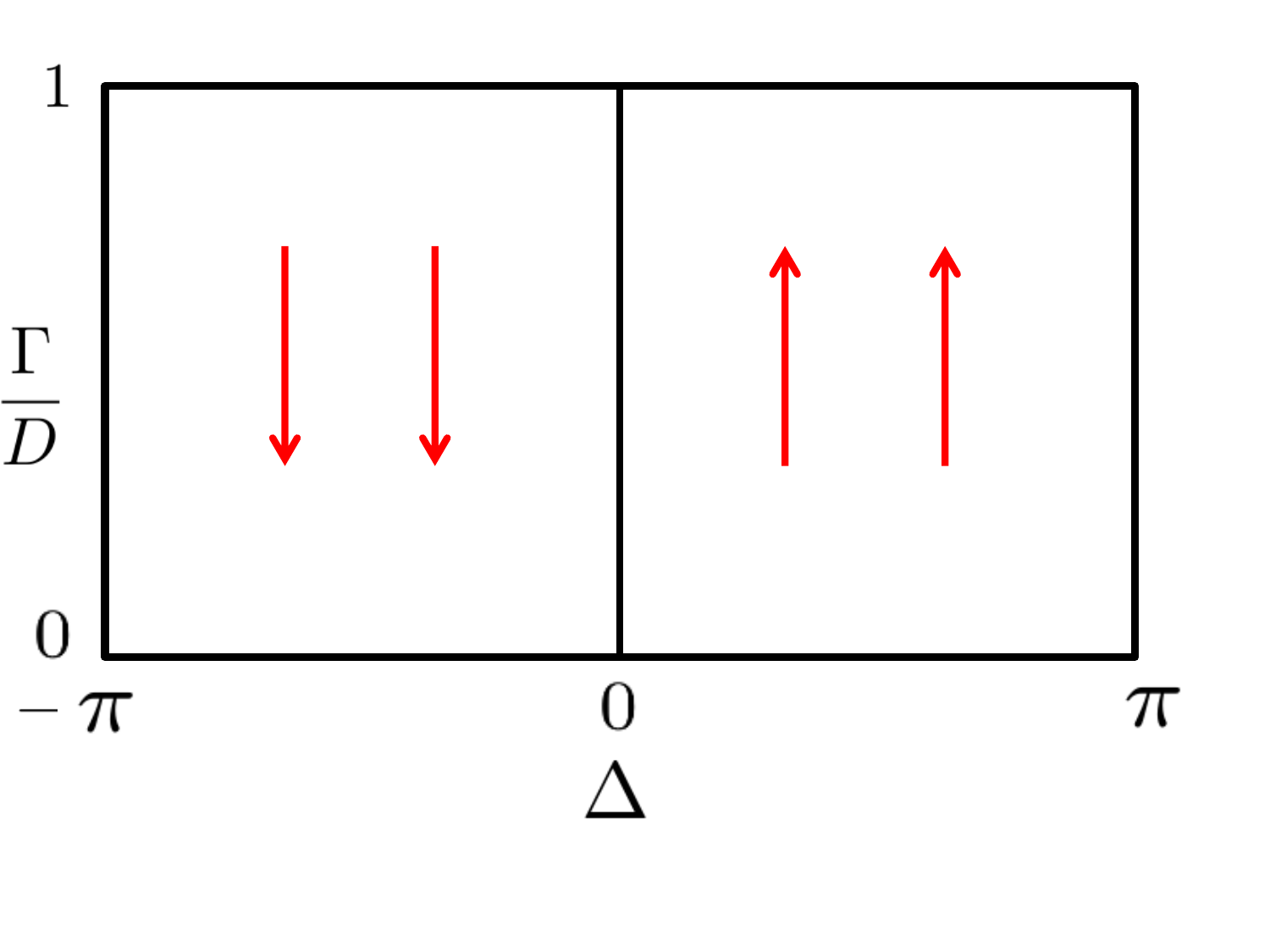}
\caption{The RG flow of the system.  For $\Delta>0$ the system flows to strong coupling and generates a scale $T_K$ allowing for universal results. For $\Delta<0$ it flows to weak coupling and the system is non universal.}
\end{figure}

We may also explore the low energy behavior of the system.  Rewriting  \eqref{Tk} as,
\begin{eqnarray}
\frac{\Gamma}{\mathcal{D}}=\gamma^{-1}\left(\frac{T_K}{\alpha\mathcal{D}}\right)^{\frac{2\Delta}{\pi}}
\end{eqnarray}
we see that reducing the cut-off \`a la Wilson, $\Gamma/\mathcal{D}$  flows to strong coupling provided $0<\Delta < \pi$.
It is interesting that despite the change in the ground state the renormalization group analysis is unaffected and so we have a unified picture for all  $0<\Delta < \pi$ of the system being weakly coupled at high energy and flowing to strong coupling at low energy. As we shall see combining results of this section with the next  the strong coupling fixed point controls the impurity behavior for low $T$ and low $\bar{\epsilon}_0$, while the weak coupling regime is reached when either of these quantities is large.

We can obtain from our expressions for the dot occupation   information about how the RG flow approaches  the strong and weak coupling fixed points by identifying the respective leading irrelevant and relevant operators \cite{ZAM2}. For $\bar{\epsilon}_0<T_K$, i.e in the region of the strong coupling fixed point the expansions for all $\Delta>0$ are given in terms of odd powers of $\bar{\epsilon}_0/T_K$ and so the leading irrelevant operator that governs the flow about the strong coupling fixed point has dimension 2  . It is natural to identify this operator with the stress energy tensor. 
 We can also extract the dimension of the leading relevant operator  around  the weak coupling fixed point i.e at high energy from the exponents in the dot occupation for $\bar{\epsilon}_0>T_K$. Again although the ground state changes form these exponents do not change and so we have the dimension of the operator is $1-\Delta/\pi$ for all $\Delta>0$. The weak coupling fixed point corresponds to the decoupled dot so the leading relevant operator is $d^\dag \psi(0)$. By setting $U=0$ we see that its dimension is $1/2K$ in agreement with perturbation theory \cite{TG} but is shifted if $U\neq 0$.

We have the following picture of the system: For $\Delta>0$ the system exhibits a renormalization group  flow from weak coupling at high energy to strong coupling at low energy. The strong coupling fixed point is at $\bar{\epsilon}_0=0$ and describes the system where the dot and the bulk are fully hybridized. By introducing an energy scale i.e. allowing  $\bar{\epsilon}_0\neq 0$ we perturb away from this fixed point. The leading irrelevant operator describing this is the stress energy tensor. The weak coupling fixed point is reached at high energy and describes a decoupled dot and bulk. By reducing the energy scale we move away from the fixed point allowing for tunneling to occur which is governed by the operator $d^\dag\psi(0)$.
 At $\Delta=0$ the system undergoes a quantum phase transition such that the low energy fixed point is no longer strongly coupled and the dot is not fully hybridized.  Any results in this regime depend upon the RG scheme used. We depict the RG flow in terms of $\Gamma/\mathcal{D}$ as a function of $\Delta\in[-\pi,\pi]$ in FIG. 4.

\section{Thermodynamic properties of the dot}

In this section we will study the system at finite temperature and calculate the free energy of the dot.  We shall find a RG flow  from weak to strong coupling as the temperature is lowered (from a localized to a delocalized dot) in agrrement to the  previously section. 
To simplify matters we specify that either $\Delta=\pi/\nu$ with $\nu>2$ being a positive integer or $\Delta=\pi-\pi/\nu$. The former covers the region $\Delta<\pi/2$ and the later $\Delta>\pi/2$. 

We begin  with $\Delta=\pi/\nu$ and also setting $\bar{\epsilon}_0=0$. For this choice of parameter there are strings of length up to $\nu-1$ (the $\nu$-string is longer present) and so excitations are created by introducing holes in these string distributions and adding particles above the Fermi sea with real rapidity. Following \cite{Takahashi} we consider the free energy $F=E-TS$ where $E$ is the energy of a state with an arbitrary configuration of strings, holes and particles,
 \begin{eqnarray}
 E=-\mathcal{D}\sum_j^{\nu-1}\int_{-\infty}^0 \frac{\sin{(j\Delta)}}{\sin{(\Delta)}}e^x\rho^p_j(x)+\mathcal{D}\int_{-\infty}^0 e^x\rho^p_+(x)
\end{eqnarray}  and $S$ is the Yang-Yang entropy 
 $S=\sum_j\int \left[(\rho^p_j+\rho_j^h)\log{(\rho^p_j+\rho_j^h)}-\rho^p_j\log{(\rho^p_j)}-\rho^h_j\log{(\rho^h_j)}\right]$ where the sum is over  $j=1,\dots,\nu-1,+$ with $+$ denoting the distributions of the real rapidity particles. We minimise $F$ with respect to $\rho_j^p$ to obtain the thermodynamic Bethe Ansatz  equations (TBA) for $\eta_j(x)\equiv \rho_j^h(x)/\rho_j^p(x)$ which determine the saddle point,
 \begin{eqnarray}\nonumber
 \log{(\eta_{j}(x))}=s\!*\!\left[\log{(1+\eta_{j-1}(x))(1+\eta_{j+1}(x))^{1+\delta_{j,\nu-2}}}\right]\\\label{TBAn}
 -\frac{2\pi}{T}\rho^0_j(x)\,.\quad\quad\quad
 \end{eqnarray}
Here $*$ denotes the convolution $f*g=\int f(x-y)g(y)\mathrm{d}y$ and additionally $ \log{(\eta_{\nu-1}(x))}=-\log{(\eta_{+}(x))}$. The driving terms of these equations, $2\pi\rho^0_j(x)$ are the energies of the fundamental excitations above the ground state, namely those obtained by adding holes to the $j$-string distributions. We can then use \eqref{TBAn} to simplify the free energy and after doing so the part which depends on the dot is given by
\begin{eqnarray}
F_d=E_d
-T\int s(x-c)\log{(1+\eta_1(x))}.
\end{eqnarray}
The first term is the ground state energy of the dot and the second term captures the finite temperature behaviour. Similar to the case of zero temperature discussed in previous sections the behaviour away from the fixed point it is determined by the $1$-string distribution. At this stage the free energy and TBA still depend on the cutoff but we can remove this dependence and take the universal limit as we did before by introducing the functions  $\varphi(x+\frac{2\Delta}{\pi}\log{\alpha' T/\mathcal{D}})$  with $1/\alpha'=\alpha\gamma^{\pi/2\Delta}$.   Taking $\mathcal{D}\to \infty$ while holding $T_K$ fixed then gives 
 \begin{eqnarray}\nonumber
 \varphi_j(x)=s*\left[\log{(1+e^{ \varphi_{j-1}(x)})(1+e^{ \varphi_{j+1}(x)})^{1+\delta_{j,\nu-2}}}\right]\\\label{UniTBA}
 -\alpha' d_je^{\frac{\pi}{2\Delta}x}\quad
 \end{eqnarray}
along with $\varphi_{\nu-1}(x)=-\varphi_+(x)$. The temperature dependent part of the free energy is now dependent on $T_K$, 
\begin{eqnarray}\label{FE}
F_d=-T\int_{-\infty}^\infty s(x+\frac{2\Delta}{\pi}\log{\left(\frac{T}{T_K}\right)})\log{(1+e^{\varphi_1(x)})}.
\end{eqnarray}
At high temperature $T\gg T_K$  the integral is dominated by $x\to-\infty$. In this limit the driving terms of \eqref{UniTBA} vanish and the solutions are given by constants $e^{\varphi_j}=(j+1)^2-1$, $e^{\varphi_{\nu-1}}=\nu-1$. Using these in the free energy (dropping the non universal part, $E_d$) we find
\begin{eqnarray}\label{FEht}
F_d(T\gg T_K)=-T\log{2}
\end{eqnarray}
which is the free energy of a two level system without energy splitting. Thus at high energy the dot is decoupled as expected from our analysis at $T=0$ of the large  $\bar{\epsilon}_0$ regime. Similarly the low temperature, $T\ll T_K$, behavior of the dot is determined by the $x\to \infty$ part of the free energy. In this case the driving terms of \eqref{UniTBA} blow up giving $\varphi_j=-\alpha'd_je^{\frac{\pi}{2\Delta}x}$ allowing us to obtain an expansion for the free energy at low temperature. We achieve this following the arguments of  \cite{TWAKM} by introducing  $\tilde{c}(\omega)=\int \exp{-i\omega x}\log(1+\exp{\varphi_1(x))}$, which is finite for Im$(\omega)>0$. Rewritten in terms of this new function the  dot free energy is 
\begin{eqnarray}\label{Fel}
F_d&=&-T\frac{1}{2\pi}\int_{-\infty}^\infty \tilde{s}(\omega)\tilde{c}(\omega)e^{-\frac{2\Delta}{\pi}i\omega\frac{T}{T_K}}\\
&=&-T\sum_{n=0}^{\infty}(-1)^n\tilde{c}(i\frac{\pi}{2\Delta}(2n+1))\left(\frac{T}{T_K}\right)^{2n+1}
\end{eqnarray}
where to obtain the second line we have closed the contour in the upper half plane and picked up the poles from $\tilde{s}(\omega)$.
The entropy of the dot $S_d=-F_d/T$ vanishes at $T=0$ as expected for a dot that is fully hybridised with the bulk.
 The coefficients of the expansion can be determined for large $n$
\begin{eqnarray}
\tilde{c}(i\frac{\pi}{2\Delta}(2n+1))=\int_{-\infty}^\infty e^{\frac{\pi}{2\Delta}(2n+1)}\log{(1+e^{\varphi_1(x)})}\\
\to \int_{-\infty}^\infty e^{\frac{\pi}{2\Delta}(2n+1)}e^{-\alpha'd_1e^{\frac{\pi}{2\Delta}x}}
=\frac{1}{(\alpha'd_1)^{2n+1}}(2n)!\,.
\end{eqnarray}
We see that the free energy is of a form similar to the expansion of the dot occupation in powers of $\bar{\epsilon}_0/T_K$  obtained at zero temperature and again the leading irrelevant operator about the strong coupling fixed point is the stress energy tensor resulting in  a power law dependence in the specific heat $C_v\sim T/T_K$.  

The thermodynamics for $\Delta=\pi-\pi/\nu$ can be investigated by similar means. We omit the details here but it can be shown that at high temperature the dot is again decoupled while at low temperature it is fully hybridized with the free energy having an expansion in terms of odd powers of $T/T_K$ as in  \eqref{Fel}.

\section{Conclusion}

We have solved via the Bethe Ansatz  the model of a Luttinger liquid coupled to an interacting resonant level at  its boundary.   We constructed the ground state and excitations of the model. It was seen that if the Luttinger interaction is sufficiently strong and repulsive (or alternatively if $U$ is strong and attractive) the ground state changes from consisting of a single type of particle to a  multicomponent condensate of strings. We then calculated the occupation of the dot as a function of the dot energy at $T=0$ obtaining exact expressions at $\Delta\ge\pi/3$ and the functional form below this. Following this we calculated the free energy of the system and studied it at low and high temperature. From these calculations we determined that  for $\Delta>0$ the system is strongly coupled at low energy and weakly coupled at high energy. The weak coupled fixed point describes a dot that is decoupled from the bulk and the leading relevant operator is the tunnelling term $\psi^\dag(0) d$, and has dimension $1-\Delta/\pi$. The strong coupling fixed point describes a fully hybridised dot and bulk with the leading irrelevant operator being the stress energy tensor.

The model has been studied  in the past in an interesting paper by Furusaki and Matveev \cite{FurMat} who identified the phase transition that occurs at $\Delta=0$ for $U=0$ below which the model is in the perturbative regime  and can be explored as an expansion in $\Gamma$ showing power law behavior of the dot occupation as a function of the dot energy.  The RG flows in this regime  are non-universal and depend on the initial values of the parameters.  We differ, however,  from these authors'  statements  (which do not bear on their main conclusions)  that the model can be mapped to the anisotropic Kondo model and that the phase transition at $\Delta=0$  corresponds to the Kondo transition from ferro- to antiferromagnetic behavior as a function of the coupling. See Appendix.

Although the model we solve here is one of a number of interacting systems with boundary impurities that are integrable (see e.g. \cite{Wang1}\cite{Wang2}\cite{Bedurftig} for some early examples) the methods we use can be extended, in conjunction with \cite{ryl} to encompass a new class of models with bulk impurities that allow for both transmission and reflection. This work will be presented elsewhere \cite{RAprl}

\acknowledgements{ We are grateful to Alexei Tsvelik, Yashar Komijani, Stefan Groha, Moshe Goldstein, Kostya Matveev,  Akira Furusaki and Hubert Saleur for useful discussions and comments. CR is supported by the Samuel Marateck Fellowship and NA by NSF Grant DMR 1410583}

\bibliography{mybib}
\begin{widetext}
\section{Appendix A}
Here we derive the  dependence of $B$  on $\bar{\epsilon}_0$ given in \eqref{B}.  Allowing for holes the energy is given by with rapidities from $-\infty$ up to an upper bound $-B$ 
\begin{eqnarray}\label{E0}
E_0/L=-\mathcal{D}\int_{-B}^{0}e^x\rho^p(x)\mathrm{d}x+\bar{\epsilon}_0\int_{-B}^0\rho^p(x)\mathrm{d}x
\end{eqnarray}
The lower limit $-B$ must be determined by minimising the  energy with respect to it for fixed $\bar{\epsilon}_0$ similar to our calculation of \eqref{comp}. We begin by inverting \eqref{bethgs} so as to find an equation for $\rho^h(x)$,
\begin{eqnarray}\label{rBAE}
\rho^0(x)=\rho^p(x)+\rho^h(x)+\int^{-B}_{-\infty}J(x-y)\rho^h(y)\mathrm{d}y.
\end{eqnarray}  
wherein $\tilde{J}=-\tilde{a}_2/(1+\tilde{a}_2)$ and the driving term is the ground state distribution from \eqref{gs 0}. 
This is a Wiener-Hopf integral and accordingly the solution is
\begin{eqnarray}\label{rhoh}
\tilde{\rho}^h(\omega)= \rho^0(-B)\frac{(G_-(\omega)G_+(i\frac{\pi}{2\Delta}))^{-1}}{\frac{\pi}{2\Delta}+i\omega}.
\end{eqnarray}
with $G_\pm(\omega)$ defined in \eqref{G}.

Inserting  \eqref{rBAE} into the expression for the energy \eqref{E0} we find that the change due to $\bar{\epsilon}_0$  is
\begin{eqnarray}
\delta E/L=\bar{\epsilon}_0\left[\int \rho^0(x)-\frac{\pi}{2(\pi-\Delta)}\int^{-B}_{-\infty}\rho^h(x)\right]
+2\pi\int_{-\infty}^{-B}\rho^0(x)\rho^h(x)
\end{eqnarray}
We recognise the first two terms as counting the number of particles minus the holes and the last term as the dressed energy of adding these holes. From \eqref{rhoh} one can see that $\rho^h(x)\propto e^{-\frac{\pi}{2\Delta}B}$ and so minimising the energy with respect to $B$ we find that
\begin{eqnarray}
e^{-\frac{\pi}{2\Delta}B}=\left(e^{\frac{\pi}{2\Delta}a}\frac{\pi(\pi-2\Delta)}{\tan{(\frac{\pi^2}{2\Delta})}}\frac{\Gamma(1+\frac{\pi}{2\Delta})}{\Gamma(\frac{1}{2}+\frac{\pi}{2\Delta})}\right)\frac{\bar{\epsilon}_0}{\mathcal{D}}.
\end{eqnarray}

We can perform an analogous calculation in the $\Delta<\pi/3$ regime where the energy is 
\begin{eqnarray}\label{Egsl}
E/L&=&-\sum_{j}^{\nu-1}\mathcal{D}\frac{\sin{(j\Delta)}}{\sin{(\Delta)}}e^{x}\rho^p_j(x)+\bar{\epsilon}_0\sum_j^{\nu-1}j\int\rho^p_j(x).
\end{eqnarray} 
By inverting the Bethe equations using \eqref{Tinv} and inserting them into \eqref{Egsl} we get shift in energy due to $\bar{\epsilon}_0$
\begin{eqnarray}
\delta E/L=\sum_j^{\nu-1}\int_{-\infty}^{-B} 2\pi\rho^0_j(x)\rho^h_j(x)+\bar{\epsilon}_0\left[\sum_{j}^{\nu-1}\int j\rho^0_j(x)-\int_{-\infty}^{-B}\frac{\pi\bar{\epsilon}_0\rho^h_{\nu-1}(x)}{2(\pi-(\nu-1)\Delta)}\right]
\end{eqnarray}
The first term is the contribution to the ground state energy due the added holes and the second and third count the number of particles minus holes. In order to minimise this we need not know the explicit form of the hole distributions but only that $\rho^h_j\propto e^{-\frac{\pi}{2\Delta}B}$ (c.f \eqref{rhoh}). Thus we find that 
\begin{eqnarray}
e^{-\frac{\pi}{2\Delta}B}=\left(\frac{1}{8(\pi-(\nu-1)\Delta)}\frac{\tilde{\rho}^h_{\nu-1}(0)}{\sum_j^{\nu-1}d_j\tilde{\rho}_j^h(-i\frac{\pi}{2\Delta})}\right)\frac{\bar{\epsilon}_0}{\mathcal{D}}.
\end{eqnarray}

\section{Appendix B}
It has been stated in the literature that the anisotropic Kondo model (AKM) is equivalent to the Luttinger dot model we study here and that further the the equivalence holds also in the absence of bulk interaction  $K=1$ and $U\neq 0$ \cite{FurMat}, \cite{gogolin2004bosonization}. In this appendix we show that in fact the AKM is not equivalent to either of these two models. 
We start by stating the AKM Hamiltonian
\begin{eqnarray}
H_{AKM}&=&\sum_{a=\uparrow,\downarrow}\int_{-L/2}^{L/2}\left(-i\psi^\dag_a \partial_x\psi_a\right)+J_z\psi^{\dag}_a(0)\psi_b(0)\sigma^z_{ab}S^z+J_\perp\left(\psi^{\dag}_a(0)\psi_b(0)\sigma^x_{ab}S^x+\psi^{\dag}_a(0)\psi_b(0)\sigma^y_{ab}S^y\right) .
\end{eqnarray}
Where the system is placed on a ring of length $L$ with periodic boundary conditions $\psi_a(x+L)=\psi_a(x)$. 
We will examine how the two models differ first using Bethe Ansatz and then via bosonization.  For the AKM the Bethe Ansatz equations for the ground state distributions of particles $\rho^p(x)$ and holes $\rho^h(x)$ are  \cite{TWAKM},
\begin{eqnarray}
D a_1(x)+\frac{1}{L}a_1(x-c')=\rho^p(x)+\rho^h(x)+\int^{\infty}_{-B'(h)}a_2(x-y)\rho^p(y)\mathrm{d}y
\end{eqnarray} 
where $\Delta$ in this context parametrizes the anisotropy of the exchange coupling, $c'$ contains the rest of the information about the impurity and $B'(h)$ the lower bound depends on the applied magnetic field, $h$. This is the ground state equation for all values $\Delta>0$. 

 We can isolate the  impurity/dot parts and compare them between the models. For the AKM this gives
\begin{eqnarray}
a_1(x-c')=\rho^p_i(x)+\rho_i^h(x)+\int^{\infty}_{-B(h)}a_2(x-y)\rho^p_i(y)\mathrm{d}y\\
\end{eqnarray} 
comparing this to \eqref{baedg}  we see that the dot part of the equations are the same however the later is valid for $\Delta\geq \pi/3$ only. As shown above, the ground state equation for the Luttinger dot model changes to \eqref{Bael} which is different from the AKM. This difference is due to the fact that the bulk of both models have different symmetries. The bulk of the AKM having $SU(2)$ symmetry which is broken to $U(1)$ by the anisotropic impurity while the Luttinger liquid only has a $U(1)$ symmetry. Therefore while the impurities in both models are similar the bulks are entirely different.

We turn now to the bosonisation of these two models and examine how they are related therein. We start with the unfolded Luttinger-dot model and take
\begin{eqnarray}
\psi(x)\sim e^{-2i\varphi(x)}
\end{eqnarray}
where $\varphi$ is a boson with the following mode expansion
\begin{eqnarray}\label{ph}
\varphi(x)=-\frac{\pi}{L}Nx-i\frac{\pi}{L}\sum \left(\frac{L|p|}{2\pi}\right)^\frac{1}{2}\frac{1}{p}e^{-ipx}\left(b^\dag_p+b_{-p}\right)e^{-|p|/2D}
\end{eqnarray}
The Hamiltonian in bosonic form is thus 
\begin{eqnarray}\label{Hirlm}
H=\frac{1}{\pi}\int_{-L/2}^{L/2}K\left(\nabla \varphi\right)^2+U'd^\dag d \nabla \varphi(0)+t'd^\dag e^{-2i\varphi(0)}+h.c
\end{eqnarray}
where we have absorbed any constants into new $U'$ and $t'$ and suppressed Klein factors. We can then absorb the Luttinger parameter into a redefinition of the field $\varphi(x)=\Phi(x)/\sqrt{K}$ to get
\begin{eqnarray}\label{Hirlm2}
H=\frac{1}{\pi}\int_{-L/2}^{L/2}\left(\nabla \Phi\right)^2+\frac{U'}{\sqrt{K}}d^\dag d \nabla \Phi(0)+t'd^\dag e^{-2i\Phi(0)/\sqrt{K}}+h.c\\\label{modek}
\Phi(x)=-\frac{\sqrt{K}\pi}{L}Nx-i\frac{\sqrt{K}\pi}{L}\sum \left(\frac{L|p|}{2\pi}\right)^\frac{1}{2}\frac{1}{p}e^{-ipx}\left(b^\dag_p+b_{-p}\right)e^{-|p|/2D}
\end{eqnarray}
Note the appearance of the factor $1/\sqrt{K}$ in the exponent of tunnelling term renders the operator therein single valued under the periodic boundary condition $x\to x+L$ also note the the change in the zero mode is reflective of the fact that the fermions are interacting.

We now perform the bosonization  of the AKM and get
\begin{eqnarray}
\frac{1}{\pi}\int_{-L/2}^{L/2}\left(\nabla \phi_\uparrow\right)^2+\left(\nabla \phi_\downarrow\right)^2+J'_z\left( \nabla\phi_\uparrow(0)-\nabla\phi_\downarrow(0) \right)+J'_\perp e^{-2i(\phi_\uparrow(0)-\phi_\downarrow(0))}S^++h.c
\end{eqnarray}where again we have the mode expansion
\begin{eqnarray}
\phi_{\uparrow,\downarrow}(x)=-\frac{\pi}{L}N_{\uparrow,\downarrow}x-i\frac{\pi}{L}\sum \left(\frac{L|p|}{2\pi}\right)^\frac{1}{2}\frac{1}{p}e^{-ipx}\left(b^\dag_{p\uparrow,\downarrow}+b_{-p,\uparrow,\downarrow}\right)e^{-|p|/2D}
\end{eqnarray}
We introduce the charge field $\phi_c=(\phi_\uparrow+\phi_\downarrow)/\sqrt{2}$ and spin field $\phi_s=(\phi_\uparrow-\phi_\downarrow)/\sqrt{2}$. These two sectors decouple and we have the spin Hamiltonian 
\begin{eqnarray}
H_s=\frac{1}{\pi}\int_{-L/2}^{L/2}\left(\nabla \phi_s\right)^2+\sqrt{2}J'_z\left( \nabla\phi_s(0)\right)+J'_\perp e^{-2\sqrt{2}i\phi_s(0)}S^++h.c
\end{eqnarray}
with 
\begin{eqnarray}\label{phis}
\phi_{s}(x)=-\frac{\pi}{\sqrt{2}L}(N_{\uparrow}-N_\downarrow)x-i\frac{\pi}{L}\sum \left(\frac{L|p|}{2\pi}\right)^\frac{1}{2}\frac{1}{p}e^{-ipx}\left(b^\dag_{p,s}+b_{-p,s}\right)e^{-|p|/2D}
\end{eqnarray}
Note that the zero mode of the spinon field has changed by a factor of $1/\sqrt{2}$ and also the Hamiltonian contains $\exp{2\sqrt{2}i\phi_s(0)}$ where as before the $\sqrt{2}$ present there is necessary for this operator to be single valued and also that the boundary conditions are correctly reproduced $\exp{2\sqrt{2}i\phi_s(0)}=\exp{2\sqrt{2}i\phi_s(L)}$. These new factors reflect the fact that $\phi_s$ is a spinon field and so does not describe a free fermion. 

Now the trick that is employed is to apply the following transformation $\mathcal{U}=\exp{(\sqrt{2}-1)S^z\phi_s(0)}$ to the Hamiltonian
\begin{eqnarray}\label{Hakm}
\mathcal{U}^\dag H_s \mathcal{U}=\frac{1}{\pi}\int_{-L/2}^{L/2}\left(\nabla \phi_s\right)^2+\sqrt{2}J''_z\left( \nabla\phi_s(0)\right)+J'_\perp e^{-2i\phi_s(0)}S^++h.c
\end{eqnarray}
The effect has been to change the coefficient in the exponent appearing in the $J'_\perp$ term back to the original one and also $J_z'\to J''_z$. Similarly one can apply the rotation  $\mathcal{U}_K=\exp{(1/\sqrt{K}-1)S^z\Phi(0)}$ to \eqref{Hirlm2}
\begin{eqnarray}\label{Hirlm3}
\mathcal{U}_K^\dag H \mathcal{U}_K=\frac{1}{\pi}\int_{-L/2}^{L/2}\left(\nabla \Phi\right)^2+\frac{U''}{\sqrt{K}}d^\dag d \nabla \Phi(0)+t'd^\dag e^{-2i\Phi(0)/\sqrt{K}}+h.c
\end{eqnarray}
 Where we have obtained a new exponent in the tunnelling term and also shifted $U'\to U''$.  At this point it is very tempting to equate \eqref{Hakm} with \eqref{Hirlm3} however while the impurity terms look the same it is important to note that for arbitrary $K$ the bulks are different as can be seen from the mode expansions of $\Phi$ and $\phi_s$. To make this more clear we can take $K=1$ in which case the bulk term of \eqref{Hirlm3} represents free fermions while that of the AKM represents spinons again this is reflected in the different zero modes of their mode expansions \eqref{phis} and \eqref{modek}. Furthermore one can note that the $e^{-2i\phi_s(0)}\neq e^{-2i\phi_s(L)}$ so the transformed AKM Hamiltonian does not respect the boundary condition. 
We can also consider the correlation function $\left<e^{-2i\phi_s(x)}e^{2i\phi_s(0)}\right>$which is no longer single valued as we can shift $x\to x+L$ in which case 
\begin{eqnarray}
\left<e^{-2i\phi_s(x)}e^{2i\phi_s(0)}\right>\to e^{-i\sqrt{2}(N_\uparrow-N_\downarrow)}\left<e^{-2i\phi_s(x)}e^{2i\phi_s(0)}\right>
\end{eqnarray}
meaning that this correlator is well defined only if $N_\uparrow-N_\downarrow=0$. We should comment that the two mode expansions agree for $K=1/2$ which corresponds to $\Delta=0$ in the Bethe language and for the AKM represents the removing the impurity while in the Luttinger-dot model is the case of maximally repulsive fermions.

Therefore in bosonisation one can also see that the two models are not equivalent. Again although the impurity parts appear the same the bulks are different.

\end{widetext}
\end{document}